\documentstyle[preprint,prd,floats,aps]{revtex}

\begin{document}
\tighten
\thispagestyle{empty}
\title{ \hfill JINR--E2--99-218 \\
  N=4 Supersymmetric Multidimensional Quantum Mechanics,
Partial SUSY Breaking and
Superconformal Quantum Mechanics
}

\author{E. E. Donets\footnote{e-mail: edonets@sunhe.jinr.ru}}
\address{Laboratory of High Energies, JINR, 141980 Dubna, Russia, }

\author{A. Pashnev\footnote{pashnev@thsun1.jinr.ru},
J. Juan Rosales\footnote{rosales@thsun1.jinr.ru} and
M. M. Tsulaia\footnote{tsulaia@thsun1.jinr.ru}}
\address{Bogoliubov Laboratory of Theoretical Physics,
 JINR, 141980 Dubna, Russia}

\maketitle

\begin{abstract}
The multidimensional $N=4$ supersymmetric quantum mechanics (SUSY QM)
is constructed using the superfield approach. As a result, the component
form of the classical and quantum Lagrangian and Hamiltonian
is obtained. In the considered SUSY QM both classical and quantum $N=4$
algebras include central charges, and this opens various possibilities
for partial supersymmetry breaking. It is shown that  quantum
mechanical models with one quarter, one half and three quarters of
 unbroken(broken) supersymmetries can exist in the framework of the
multidimensional $N=4$ SUSY QM, while the one-dimensional $N=4$ SUSY QM,
constructed earlier, admits only  one half or total supersymmetry
breakdown. We illustrate the constructed general formalism, as well as
all possible cases of the partial SUSY breaking taking as an example
a direct multidimensional generalization of the one-dimensional
$N=4$ superconformal quantum mechanical model. Some open questions and
possible applications of the constructed multidimensional $N=4$ SUSY QM
to the known exactly integrable systems and  problems of quantum
cosmology are briefly discussed. \\
\end{abstract}

\newpage
\section{INTRODUCTION}

The supersymmetric quantum mechanics (SUSY QM), being first introduced
in  \cite{WW} -- \cite{ED} for the $N=2$ case, turns out to be a
convenient
tool for investigating problems of  supersymmetric field theories,
since it provides the simple and, at the same time, quite adequate
understanding of various phenomena arising in  relativistic theories.

The important question of all modern theories of fundamental interactions,
including superstrings and M~ -- theory,
is the problem of  spontaneous breakdown of supersymmetry.
Supersymmetry, as a fundamental symmetry of the nature, if exists,
has to be spontaneously broken at low energies since particles with
all equal quantum numbers, except the spin, are not observed
experimentally.
 Several (rather different) mechanisms of  spontaneous breakdown of
supersymmetry have been proposed in particle physics in order to resolve
this problem. One of them is to add to the supersymmetric Lagrangian
the so -- called $D$~--, or $F$~-- terms which are invariant under
supersymmetry
transformations  but break  supersymmetry spontaneously due to
 nonzero vacuum expectation values or, alternatively,
introduce into the theory some
 soft breaking mass terms ''by hand"; the latter procedure does not spoil
the nonrenormalization theorem of the supersymmetric field
theories and was successfully applied to construct the Minimal
Supersymmetric extension of the Standard Model (see  \cite{SM}
and Refs. therein).  Next mechanism of SUSY breaking is the
dynamical (nonperturbative) breakdown of  supersymmetry, caused
by instantons (see, for example, \cite{SWZ} and Refs. therein). In
this case, the energy of tunneling between topologically
distinct vacua produces an energy shift from the zero level,
hence leading to the spontaneous breakdown of supersymmetry. And,
finally, the mechanism of  partial spontaneous breaking of the
$N=2$ supersymmetry in the field theory was recently proposed  in
 \cite{ANTONIAD}. This mechanism is based on the inclusion
into the Lagrangian two types of the Fayet -- Iliopoulos terms --
electric and magnetic ones, and it leads to the corresponding
modification of the $N=2$ SUSY algebra  \cite{IZ}.

The problem of  spontaneous breakdown of supersymmetry
could be investigated in the framework of the supersymmetric quantum
mechanics as well. The conjecture that supersymmetry
can be spontaneously broken by instantons \cite{WW} -- \cite{ED}
was investigated in detail by several authors for the case
of $N=2$ SUSY QM \cite{SH} -- \cite{KM}.
However, the most physically interesting case
is provided by the $N=4$ supersymmetric
quantum mechanics since it can
be applied to the  description of the systems resulting from the
''realistic" $N=1$ supersymmetric field theories (including supergravity)
in four ($D=4$) dimensions after the dimensional reduction to one
dimension (see, for example, \cite{CH}).

The one -- dimensional $N=4$ SUSY QM was constructed first in
\cite{BP1} -- \cite{IKP}.  Partial breaking of supersymmetry, caused by
the presence of the
central charges in the corresponding superalgebra, was
also discussed in  \cite{IKP}. It was the first example of  partial
breaking of supersymmetry in the framework of SUSY QM and the corresponding
mechanism is in full analogy with that in  \cite{ANTONIAD} in the field
theory. The main point is that the presence
of  central charges in the superalgebra allows
the partial supersymmetry breakdown, whereas according to
Witten's theorem \cite{WW}, no partial supersymmetry breakdown is
possible if the SUSY algebra includes no central charges.
The main goal of our paper is further generalization of the construction,
proposed in  \cite{IKP} to the multidimensional case and
investigation of
partial breaking of supersymmetry under  consideration.

 Consideration of the supersymmetric algebra with
central charges is of particular importance  for  several reasons.
 First, it provides a good tool to
study  dyon solutions of quantum field theory since in such theories
the mass and electric and magnetic charges  turns out
to be the central charges \cite{WO}. Second,  the central
charges produce the rich structure of supersymmetry breaking. Namely,
it is possible to break  part of all supersymmetries retaining all others
exact \cite{WESS}.
In fact, the invariance of a state with respect to the supersymmetry
transformation means  saturation of the Bogomol'ny bound, and this
situation takes place in $N=2$ and $N=4$ supersymmetric Yang -- Mills
theory \cite{GAU} -- \cite{BLUM} as well as in the theories of  extended
supergravity \cite{KLO}.

 The investigation of
supersymmetric properties of branes the in M~ -- theory
has also revealed that
 partial breakdown of supersymmetry takes place. Namely, the ordinary
branes break half of the supersymmetries, while ``intersecting" and
rotating
branes can leave only $1/4$,  $1/8$, $1/16$ or $1/32$ of the
supersymmetries unbroken
\cite{STT}.

The main characteristic features of   partial SUSY breaking
in the field theories with the extended supersymmetry can be revealed
in supersymmetric QM, since in the both cases  partial supersymmetry
breakdown is provided by the central charges in the SUSY algebra.
Therefore the detailed study of  partial supersymmetry
breakdown in supersymmetric quantum mechanics can lead to the deeper
understanding of analogous effect in supersymmetric field theories.

 The known examples of the breakdown of  supersymmetries in the
supersymmetric quantum mechanics are the cases, where either all
supersymmetries are broken(exact) or only  half of them
are broken(exact) \cite{IKP}. In this paper, we demonstrate
the possibility of the  three -- quarters or one -- quarter of
the supersymmetry breakdown in the framework of the multidimensional
$N=4$ SUSY QM.
The later case ($3/4$ of the supersymmetries are exact) has not been
observed  before
 in SUSY QM
( in the specific $N=4$ supergravity model it was observed in
\cite{TZ}) and
seems to be quite interesting by itself even without specifying the
physical origin of the phenomenon.

The paper is organized as follows. In Sec. II, we present a formal
construction of the $N=4$ multidimensional supersymmetric quantum
mechanics: classical and quantum Hamiltonian and Lagrangian, SUSY
transformations, algebra of supercharges  and so on. In Sec. III,
partial supersymmetry breaking is investigated and all possible
cases of the partial SUSY breakdown are listed. In Sec. IV, we give
an exactly solvable example which illustrates main properties of
the introduced formal constructions. This example is interesting
by itself since we consider the multidimensional generalization of
the $N=4$ superconformal quantum mechanics \cite{IKP}, \cite{IKL}
which is naturally related to the extremal RN black holes in
''near horizon" limit and adS/CFT correspondence \cite{KALLOSH}.
In Sec. V, we conclude with some open questions and further
perspectives. \\

\setcounter{equation}0\section{D -- dimensional $N=4$ SUSY Quantum
Mechanics}

In this section, we describe a general formalism of
the $D$~-- dimensional
($D \ge 1$) $N=4$ supersymmetric
quantum mechanics, starting with the superfield approach
and concluding with the component form of the desired Lagrangian and
Hamiltonian.

 Consider $N=4$ SUSY transformations
\begin{eqnarray} \nonumber
\delta t &=& \frac{i}{2} (\epsilon^a \bar \theta_a + \bar
\epsilon_a \theta^a) , \nonumber \\ \delta \bar \theta_a &=& \bar
\epsilon_a \nonumber , \\ \delta \theta^a &=&\epsilon^a ,
\end{eqnarray}
in the superspace spanned by the even coordinate  $t$ and
mutually complex -- conjugated odd coordinates
$\theta^a$ and $\bar \theta_a$. The parameters of $N=4$
SUSY transformations $\epsilon^a$ and $\bar \epsilon_a$ are complex
conjugate to each other as well.\footnote{Our conventions for spinors are
as follows:
${\theta}_{a}
={\theta}^{b}{\varepsilon}_{ba},\; {\theta}^{a}=
{\varepsilon}^{ab}{\theta}_{b},\;
{\bar \theta}_{a}={\bar \theta}^{b}{\varepsilon}_{ba},\; {\bar \theta}^{a}=
{\varepsilon}^{ab}{\bar \theta}_{b},\;
{\bar \theta}_a = (\theta^a)^\ast,\;
{\bar \theta}^a = -(\theta_a)^\ast,\;
(\theta \theta)\equiv \theta^a \theta_a = -2 \theta^1 \theta^2,\;
(\bar \theta \bar \theta)\equiv \bar \theta_a
\bar \theta^a = (\theta \theta)^\ast,\;
\varepsilon^{12} = 1,\;
\varepsilon_{12} = 1$.}
The generators of the above supersymmetry transformations
\begin{equation}
Q_a = \frac{\partial}{\partial \theta^a} + \frac{i}{2} \bar \theta_a
\frac{\partial}{\partial t} ,
\quad
\bar Q^ a = \frac{\partial}{\partial \bar \theta_a} + \frac{i}{2} \theta^a
\frac{\partial}{\partial t} ,
\end{equation}
along ~with ~the ~time ~translation ~operator $H = i
\frac{\partial}{\partial t}$ ~obey ~the ~following
~(anti)commutation ~relations: $$ \{ Q_a, \bar Q^b \}= \delta^b_a
H, $$
\begin{equation} \label{XXX}
[H, Q_a] = [H, \bar Q^a]= 0 .
\end{equation}
The automorphism group for a given algebra is $SO(4) = SU(2) \times SU(2)$
and the generators of the $N=4$ SUSY transformations
are in the spinor representation of one of the $SU(2)$ groups.

The next step is to construct irreducible representations
of the algebra (\ref{XXX}). The usual way of doing this is to use the
supercovariant derivatives
\begin{equation}
D_a = \frac{\partial}{\partial \theta^a}- \frac{i}{2} \bar \theta_a
\frac{\partial}{\partial t} ,
\quad
\bar D^a = \frac{\partial}{\partial \bar \theta_a}- \frac{i}{2} \theta^a
\frac{\partial}{\partial t} ,
\end{equation}
and impose some constraints on the general superfield. Hereafter we deal
with the superfield $\Phi^i$ $(i=1,...,D)$ subjected to
the following constraints:
\begin{eqnarray} \nonumber
[ D_a , \bar D^a]\Phi^i &=& -4m^i , \\ \nonumber
D^a D_a \Phi^i &=& -2n^i , \\
\bar D_a \bar D^a \Phi^i &=& -2\bar n^i ,
\end{eqnarray}
where $m^i$ are real constants, while $n^i$ and $\bar n^i$ are
mutually complex -- conjugated constants. Such constraints for the
case of the one -- dimensional N =4 SUSY QM
 were considered first in \cite{IKP} as a minimal
consistent generalization of the analogous constraints of
  \cite{BP1}  and   \cite{IKL}  with  vanishing right hand side.
The presence of the additional arbitrary  parameters leads, as we shall
see below to the considerably richer structure of the theory.
 The explicit form of the
superfield $\Phi^i$ is the following:
\begin{eqnarray} \nonumber
\Phi^i&=&\phi^i + \theta^a \bar \psi^i_a - \bar \theta_a
\psi^{ia} + \theta^a B^{bi}_a \bar \theta_b + \\ \nonumber &+&
m^i ( \theta \bar \theta)  +\frac{1}{2}n^i( \theta  \theta)   +
\frac{1}{2}\bar n^i (\bar \theta \bar \theta) +  \\ \label{SF}
&+&\frac{i}{4}(\theta \theta) \bar \theta_a \dot{\bar \psi^{ai}}-
\frac{i}{4}(\bar\theta \bar\theta) \theta^a \dot{\psi_a^i} +
\frac{1}{16}(\theta \theta)(\bar \theta \bar \theta)
\stackrel{..}{\phi^i} ,
\end{eqnarray}
($\hspace{0.2cm} \dot{} \equiv \partial_t$). Note that in the case when
all the constants $m^i$, $n^i$ and $\bar n^i$ are
equal to zero, the superfield   (\ref{SF}) represents $D$ ''trivial"
copies of the superfield $\Phi$, given in \cite{BP1}, which describes
the
irreducible representation of the one-dimensional $N=4$ SUSY QM.
The latter superfield contains one
bosonic field $\phi$, four fermionic fields
$\psi^a$ and $\bar \psi_a$, and three auxiliary bosonic fields
$B^b_a = (\sigma_I)^b_a B^I$ where $(\sigma_I)^b_a$
$(I = 1,2,3)$ are ordinary Pauli matrices.

Another irreducible representation of the algebra (\ref{XXX})
can be constructed after making the appropriate generalization
of the constraints given in  \cite{BP2}:
\begin{equation}
({\varepsilon}^{ac}D_c \bar D^b  +
{\varepsilon}^{bc}D_c \bar D^a)\Phi =0 .
\end{equation}
The technique of constructing $N=4$ SUSY invariant Lagrangians
is absolutely the same for both the cases and, therefore, we shall not
consider the second one separately.

The components of the superfield (\ref{SF})
 transform under the $N=4$ transformations as follows:
\begin{eqnarray} \nonumber
\delta \phi^i&=&\epsilon^a \bar \psi^i_a - \bar \epsilon_a  \psi^{ai} , \\
\nonumber
\delta \psi^{ai}&=&\epsilon^b B_b^{ai} + \frac{i}{2}\epsilon^a
\dot{\phi^i} + \epsilon^a m^i - \bar \epsilon^a \bar n^i ,
\\ \label{SUSYT}
\delta B_{b}^{ai}&=& -\frac{i}{2} \epsilon_b \dot {\bar \psi^{ai}}
-\frac{i}{2} \epsilon^a \dot{\bar \psi_b^{i}}
-\frac{i}{2} \bar \epsilon^a \dot{ \psi_b^{i}}
-\frac{i}{2} \bar \epsilon_b \dot{ \psi^{ai}} .
\end{eqnarray}

Now one can write down the most general form of the Lagrangian which is
invariant under the above-mentioned $N=4$ SUSY transformations
\begin{equation} \label{LAGR}
L=-8( \int d^2\theta d^2 \bar \theta (A(\Phi^i)) +
\frac{1}{16}\lambda^{a}_{bi} B^{bi}_{a}) ,
\end{equation}
where $A(\Phi^i)$ is an arbitrary function of the superfield
$\Phi^i$ called the superpotential. The second term is the
{}Fayet -- Iliopoulos term and
$\lambda^{a}_{bi}=(\sigma_I)^a_b \Lambda^I_i $ are just constants.
The expression for the Lagrangian (\ref{LAGR}) is the most general one
in the sense that any other $N=4$ SUSY invariant terms added will
lead with necessity to  higher derivatives in the component form.

After the integration with respect to the Grassmanian coordinates
$\theta^a$ and $\bar \theta_a$, one obtains the component form of the
Lagrangian (\ref{LAGR})
\begin{equation} \label{COMP}
L = K - V ,
\end{equation}
where
\begin{equation}
K = \frac{1}{2}\frac{\partial^2 A}{\partial \phi^i \partial
\phi^j}\dot \phi^i
\dot \phi^j + i \frac{\partial^2 A}{\partial \phi^i \partial \phi^j}
(\bar \psi_a^i \dot \psi^{aj} + \psi^{ai} \dot{\bar \psi_a^j}) ,
\end{equation}
and
\begin{eqnarray} \nonumber
V&=&2\frac{\partial^2 A}{\partial \phi^i \partial \phi^j}
(m^im^j +  n^i\bar n^j)
-\frac{\partial^2 A}{\partial \phi^i \partial \phi^j}
B^{ai}_b B^{bj}_a +
\\ \nonumber
&+& \frac{\partial^3 A}{\partial \phi^i \partial \phi^j \partial \phi^k}
(2\bar \psi^i_a \psi^{aj} m^k + \psi^{ai} \psi_{a}^{j} n^k +
\bar \psi^{i}_{a} \bar \psi^{aj} \bar n^k)
-
\frac{\partial^3 A}{\partial \phi^i \partial \phi^j \partial \phi^p}
(\bar \psi^i_a \psi^{bj} + \bar \psi^{ib} \psi_{a}^j)B^{ap}_b +
 \\
&+&\frac{1}{2}
\frac{\partial^4 A}{\partial \phi^i \partial \phi^j \partial
\phi^k \partial
\phi^l}
(\bar \psi^i_a \bar \psi^{ja})(\psi^{bk} \psi_b^l)
+\frac{1}{2} \lambda^a_{bi}B^{bi}_a   .
\end{eqnarray}
Expressing the auxiliary field $B^{bi}_a$ in terms of the physical
fields
\begin{equation}
B^{ai}_b =
{(\frac{\partial^2 A}{\partial \phi^i \partial \phi^j})}^{-1}
(\frac{1}{4} \lambda^{a}_{bj} - \frac{1}{2}
\frac{\partial^3 A}{\partial \phi^j \partial \phi^k \partial \phi^p}
(\bar \psi^k_a \psi^{bp} + \bar \psi^{bk} \psi_{a}^p)) ,
\end{equation}
using its equation of motion
and inserting it back into the Lagrangian
(\ref{COMP}), one obtains a final
form of the potential term
\begin{eqnarray} \nonumber
V&=&\frac{1}{16} \lambda^a_{bi}\lambda^{b}_{aj}
{(\frac{\partial^2 A}{\partial \phi^i \partial \phi^j})}^{-1}  +
2\frac{\partial^2 A}{\partial \phi^i \partial \phi^j}
(m^im^j +  n^i\bar n^j) + \\ \nonumber
&+& \frac{\partial^3 A}{\partial \phi^i \partial \phi^j \partial \phi^k}
(2\bar \psi^i_a \psi^{aj} m^k + \psi^{ai} \psi_{a}^{j} n^k +
\bar \psi^{i}_{a} \bar \psi^{aj} \bar n^k) - \\ \nonumber
&-& \frac{1}{2} \lambda^a_{bp}{(\frac{\partial^2 A}{\partial \phi^p
\partial \phi^k})}^{-1}
\frac{\partial^3 A}{\partial \phi^i \partial \phi^j \partial \phi^k}
\bar \psi^i_a  \psi^{bj} +  \\ \nonumber
&+&\frac{1}{2}
\frac{\partial^4 A}{\partial \phi^i \partial \phi^j \partial \phi^k
\partial \phi^l}
(\bar \psi^i_a \bar \psi^{ja})(\psi^{bk} \psi_b^l)
-{(\frac{\partial^2 A}{\partial \phi^p \partial \phi^q})}^{-1}
(\frac{\partial^3 A}{\partial \phi^i \partial \phi^k \partial \phi^p}
\frac{\partial^3 A}{\partial \phi^q \partial \phi^j\partial \phi^l} + \\
&+& \frac{1}{2}
\frac{\partial^3 A}{\partial \phi^i \partial \phi^j \partial \phi^p}
\frac{\partial^3 A}{\partial \phi^q \partial \phi^k\partial \phi^l})
\bar \psi^j_{a} \bar \psi_b^k \psi^{bl} \psi^{ai} ,
\end{eqnarray}
where  the identity
$$
{(\frac{\partial^2 A}{\partial \phi^p \partial \phi^q})}^{-1}
\frac{\partial^3 A}{\partial \phi^i \partial \phi^k \partial \phi^p}
\frac{\partial^3 A}{\partial \phi^q \partial \phi^j\partial \phi^l}
\bar \psi^i_a \bar \psi^{al} \psi^{jb} \psi^k_b
=
$$
\begin{equation}
={(\frac{\partial^2 A}{\partial \phi^p \partial \phi^q})}^{-1}
\frac{\partial^3 A}{\partial \phi^i \partial \phi^k \partial \phi^p}
\frac{\partial^3 A}{\partial \phi^q \partial \phi^j\partial \phi^l}
(\bar \psi^i_a  \psi^{aj} \bar \psi_b^l \psi^{bk} +
\bar \psi^i_a  \psi^{ak} \bar \psi_b^l \psi^{bj})
\end{equation}
was used.

The formulae given above, can be rewritten in a different and more
natural form using the geometrical notation. Let us introduce
the metric of some ``target" manifold in the following way:
\begin{equation} \label{METR}
g_{ij}= \frac{\partial^2 A}{\partial \phi^i \partial \phi^j} ,
\end{equation}
along with the corresponding Christoffel connection and the Riemann
curvature
\begin{equation}  \label{CHRIST}
\Gamma^i_{jk}= \frac{1}{2}
\frac{\partial^3 A}{\partial \phi^p \partial \phi^j \partial \phi^k}
{(\frac{\partial^2 A}{\partial \phi^p \partial \phi^i})}^{-1} ,
\end{equation}
\begin{equation}
R_{ij,kl}= \frac{1}{4}
{(\frac{\partial^2 A}{\partial \phi^p \partial \phi^q})}^{-1}
(\frac{\partial^3 A}{\partial \phi^i \partial \phi^l \partial \phi^p}
\frac{\partial^3 A}{\partial \phi^q \partial \phi^j \partial \phi^k}
-
\frac{\partial^3 A}{\partial \phi^i \partial \phi^k \partial \phi^p}
\frac{\partial^3 A}{\partial \phi^q \partial \phi^j \partial \phi^l}) .
\end{equation}
Now the Lagrangian (\ref{COMP}),  rewritten in terms of these
geometric quantities looks as follows:
\begin{equation}
K = \frac{1}{2}g_{ij}
\dot \phi^i \dot \phi^j + i g_{ij}
(\bar \psi_a^i \dot \psi^{aj} + \psi^{ai} \dot{\bar \psi_a^j}) ,
\end{equation}
and
\begin{eqnarray}  \nonumber
V&=&\frac{1}{16} \lambda^a_{bi}\lambda^{b}_{aj}
g^{ij}  + 2 g_{ij}
(m^im^j +  n^i\bar n^j) + \\ \nonumber
&+&4\bar \psi^i_a \psi_j^{a}D_i m^j  + 2\psi^{ai} \psi_{aj} D_i n^j +
2\bar \psi^{i}_{a} \bar
\psi_j^{a} D_i\bar n^j  +
\bar \psi^i_a  \psi^{bj} D_i\lambda^a_{bj} + \nonumber \\
&+& (D_{i}\Gamma_{jkl} + R_{ik,lj})
\bar \psi^i_{a} \bar \psi^{aj} \psi^{bk} \psi_b^{l}
+ R_{jl,ki}
\bar \psi^i_{a}  \psi^{aj} \bar \psi_b^k \psi^{bl} ,
\end{eqnarray}
where $D_i$ is a standard covariant derivative defined with the help of
the
introduced Christoffel connection (\ref{CHRIST}).
 Using the Noether theorem technique, one can find
the classical expressions for the conserved supercharges,
corresponding to the SUSY transformations (\ref{SUSYT})
\begin{eqnarray} \nonumber
\bar Q_a &=& \bar\psi^i_a p_i - 2i \bar\psi^i_a m^j
\frac{\partial^2 A}{\partial \phi^i \partial \phi^j}
+2i \psi_a^i n^j \frac{\partial^2 A}{\partial \phi^i \partial \phi^j} +
\\ \label{CHARGE1}
&+&\frac{i}{2} \bar \psi^i_c \bar \psi^{cj}  \psi_a^k
\frac{\partial^3 A}{\partial \phi^i \partial \phi^j \partial \phi^k}
- \frac{1}{2}i \lambda^c_{ai} \bar \psi^i_c ,
\end{eqnarray}

\begin{eqnarray}  \nonumber
Q^b&=&\psi^{ib} p_i + 2i \psi^{bi} m^j
\frac{\partial^2 A}{\partial \phi^i \partial \phi^j}
+2i \bar \psi^{bi} \bar n^j
\frac{\partial^2 A}{\partial \phi^i \partial \phi^j} + \\ \label{CHARGE2}
&+&\frac{i}{2} \bar \psi^{bi}  \psi^{cj}  \psi_c^k
\frac{\partial^3 A}{\partial \phi^i \partial \phi^j \partial \phi^k}  +
\frac{1}{2}i \lambda^b_{di} \psi^{di} .
\end{eqnarray}
These formulae for the conserved supercharges complete the classical
description of the desired $N=4$ SUSY multidimensional mechanics,
and now to quantize it we should analyze its constraints.

{}Following the standard procedure of quantization of the system with
bosonic and fermionic degrees of freedom \cite{RC},
we introduce the canonical Poisson brackets
\begin{equation}
\{ \phi^i , p_{j} \}=\delta^i_j , \quad
\{ \psi^{ai} , p_{(\psi), bj} \}=-\delta^a_b\delta^i_j ,
\quad \{ \bar \psi_a^i , p^b_{(\bar\psi),j} \}=-\delta^a_b\delta^i_j ,
\end{equation}
 where $p_{i},p_{(\psi), ai}$, and $p^a_{(\bar\psi),i}$ are the
momenta conjugated to $\phi^i,\psi^{ai}$ and $\bar \psi_a^i$.
{}From the explicit form of the momenta
\begin{equation}
p_i= g_{ij}{\dot{\phi}}^i ,
\end{equation}
\begin{equation}
p_{(\psi), ai}=-ig_{ij}\bar \psi_a^j , \quad
p^a_{(\bar \psi), i}=-ig_{ij} \psi^{aj} ,
\end{equation}
with the  metric $g_{ij}$ given by (\ref{METR}),
one can conclude  that the system possesses the
second -- class fermionic constraints
\begin{equation}
\chi_{(\psi), ai} =  p_{(\psi), ai} + ig_{ij}\bar \psi_a^j ,
\quad \hbox{and}
\quad \chi^a_{(\bar \psi), i} = p^a_{(\bar \psi), i} + ig_{ij} \psi^{aj} ,
\end{equation}
 since
\begin{equation}
\{ \chi^a_{(\bar \psi), i} , \chi_{(\psi), bj} \} = -2i g_{ij}\delta^a_b .
\end{equation}
Therefore, the quantization has to be done using the
Dirac brackets defined for any  two functions $V_a$ and $V_b$ as
\begin{equation} \label{DR}
\{V_a , V_b \}_{Dirac} =
\{V_a , V_b \} - \{V_a , \chi_c \} \frac{1}{\{\chi_c , \chi_d \}}
\{\chi_d , V_b \} .
\end{equation}
 As a result, we obtain the
following Dirac brackets for the canonical variables:
$$
\{ \phi^i, p_j \}_{Dirac} = \delta^i_j ,
\quad
\{ \psi^{ai}, \bar \psi^j_b \}_{Dirac} = - \frac{i}{2} \delta^a_b
{(\frac{\partial^2 A}{\partial \phi^i \partial \phi^j})}^{-1} =
- \frac{i}{2} \delta^a_b g^{ij} ,
$$
\begin{eqnarray} \nonumber
\{ \psi^{ai}, p_j \}_{Dirac}&=&- \frac{1}{2}  \psi^{ap}
\frac{\partial^3 A}{\partial \phi^p \partial \phi^m \partial \phi^j}
{(\frac{\partial^2 A}{\partial \phi^m \partial \phi^i})}^{-1}
= - \psi^{ak} \Gamma^i_{jk} , \\
\{ \bar \psi^i_a, p_j \}_{Dirac}&=&- \frac{1}{2}  \bar \psi^p_a
\frac{\partial^3 A}{\partial \phi^p \partial \phi^m \partial \phi^j}
{(\frac{\partial^2 A}{\partial \phi^m \partial \phi^i})}^{-1} =
- \bar \psi^k_a  \Gamma^i_{jk} ,
\end{eqnarray}
and, finally,
\begin{eqnarray} \nonumber
\{ p_i, p_j \}_{Dirac}&=&- \frac{i}{2}
{(\frac{\partial^2 A}{\partial \phi^p \partial \phi^q})}^{-1}
(\frac{\partial^3 A}{\partial \phi^i \partial \phi^k \partial \phi^p}
\frac{\partial^3 A}{\partial \phi^q \partial \phi^j \partial \phi^l}-
\frac{\partial^3 A}{\partial \phi^i \partial \phi^l \partial \phi^p}
\frac{\partial^3 A}{\partial \phi^q \partial \phi^j \partial \phi^k})
\bar \psi_a^k \psi^{al} \\
&=&2i R_{ij,kl}\bar \psi_a^k \psi^{al} .
\end{eqnarray}

The classical Hamiltonian, obtained after the usual Legendre
transformation from the Lagrangian (\ref{COMP}), has the form
\begin{equation} \label{HAM}
H_{class.}= \frac{1}{2}
{(\frac{\partial^2 A}{\partial \phi^i \partial \phi^j})}^{-1}p_i p_j + V .
\end{equation}

The supercharges and the Hamiltonian form the following $N=4$
SUSY algebra with respect to the introduced Dirac brackets
$$
\{ \bar Q_a , Q^b \}_{Dirac} = -i \delta^b_a H_{class.} -
i \lambda^b_{ai}m^i ,
$$
\begin{equation}
\{ \bar Q_a , \bar Q_b \}_{Dirac} =  - i \lambda_{abi}n^i , \quad
\{  Q^a , Q^b \}_{Dirac} =  i \lambda^{ab}_i \bar n^i .
\end{equation}
Note the appearance of the central charges in the algebra.
This fact is extremely important especially for the investigation
of partial supersymmetry breaking, given in the next section.

Replacing the Dirac brackets by (anti)commutators using the rule
\begin{equation}
i\{ , \}_{Dirac} =   \{ , \} ,
\end{equation}
one obtains the quantum algebra
$$
\{ \bar Q_a , Q^b \} =  \delta^b_a H_{quant.} +   \lambda^b_{ai}m^i ,
$$
\begin{equation} \label{OOO}
\{ \bar Q_a , \bar Q_b \} =   \lambda_{abi}n^i , \quad
\{  Q^a , Q^b \} =  -  \lambda^{ab}_i \bar n^i ,
\end{equation}
under  the definite choice of operator ordering in the
supercharges (\ref{CHARGE1}) -- (\ref{CHARGE2}) and in the Hamiltonian
(\ref{HAM})
\begin{equation}  \label{QUANT1}
\bar Q_a=\bar \psi^i_a R_i  - 2i \bar \psi^i_a m^j
\frac{\partial^2 A}{\partial \phi^i \partial \phi^j}
+2i \psi_a^i n^j \frac{\partial^2 A}{\partial \phi^i \partial \phi^j}
- \frac{1}{2}i \lambda^c_{ai} \bar \psi^i_c ,
\end{equation}
\begin{equation}
Q^b=L_i\psi^{bi}  + 2i \psi^{bi} m^j \label{QUANT2}
\frac{\partial^2 A}{\partial \phi^i \partial \phi^j}
+2i \bar \psi^{bi} \bar n^j
\frac{\partial^2 A}{\partial \phi^i \partial \phi^j}
+\frac{1}{2}i \lambda^b_{di} \psi^{di} ,
\end{equation}
\begin{eqnarray}  \nonumber
H_{quant.}&=&\frac{1}{2}L_i
{(\frac{\partial^2 A}{\partial \phi^i \partial \phi^j})}^{-1}R_j
+ \frac{1}{16}\lambda^a_{bi}\lambda^{b}_{aj}
{(\frac{\partial^2 A}{\partial \phi^i \partial \phi^j})}^{-1}  +
2\frac{\partial^2 A}{\partial \phi^i \partial \phi^j}
(m^im^j +  n^i\bar n^j) + \\ \nonumber
&+& \frac{\partial^3 A}{\partial \phi^i \partial \phi^j \partial \phi^k}
( [\bar \psi^i_a \psi^{aj}] m^k + \psi^{ai} \psi_{a}^{j} n^k +
\bar \psi^{i}_{a} \bar \psi^{aj} \bar n^k) - \\ \nonumber
&-&\frac{1}{4} \lambda^a_{bp}{(\frac{\partial^2 A}
{\partial \phi^p \partial \phi^k})}^{-1}
\frac{\partial^3 A}{\partial \phi^i \partial \phi^j \partial \phi^k}
[\bar \psi^i_a , \psi^{bj}]
+\frac{1}{2}
\frac{\partial^4 A}{\partial \phi^i \partial \phi^j \partial \phi^k
\partial \phi^l}
(\bar \psi^i_a \bar \psi^{ja})(\psi^{bk} \psi_b^l) -  \\ \nonumber
&-&{(\frac{\partial^2 A}{\partial \phi^p \partial \phi^q})}^{-1}
(\frac{\partial^3 A}{\partial \phi^i \partial \phi^k \partial \phi^p}
\frac{\partial^3 A}{\partial \phi^q \partial \phi^j\partial \phi^l} + \\
\label{OPERATOR}
&+& \frac{\partial^3 A}{\partial \phi^i \partial \phi^j \partial \phi^p}
\frac{\partial^3 A}{\partial \phi^q \partial \phi^k\partial \phi^l})
\bar \psi^j_{a} \bar \psi_b^k \psi^{bl} \psi^{ai} ,
\end{eqnarray}
where
\begin{eqnarray} \nonumber
L_i&=&p_i + i \bar \psi_a^j \psi^{ak}
\frac{\partial^3 A}{\partial \phi^i \partial \phi^j \partial \phi^k}
-\frac{i}{2}
{(\frac{\partial^2 A}{\partial \phi^j \partial \phi^k})}^{-1}
\frac{\partial^3 A}{\partial \phi^i \partial \phi^j \partial \phi^k} , \\
R_i&=&p_i - i \bar \psi_a^j \psi^{ak}
\frac{\partial^3 A}{\partial \phi^i \partial \phi^j \partial \phi^k}
+\frac{i}{2}
{(\frac{\partial^2 A}{\partial \phi^j \partial \phi^k})}^{-1}
\frac{\partial^3 A}{\partial \phi^i \partial \phi^j \partial \phi^k} .
\end{eqnarray}
~The ~momentum ~operators ~are ~Hermitean ~with ~respect ~to
~the ~integration ~measure
$d^D \phi \sqrt{| det{(\frac{\partial^2 A}
{\partial \phi^i \partial \phi^j})}^{-1}|}$
if they have the following
form:
\begin{equation} \label{IMP}
p_i = -i \frac{\partial}{\partial \phi^i} -
\frac{i}{4} \frac{\partial}{\partial \phi^i} ln (| det g_{ik}|)
-2i \omega_{i \alpha \beta}\bar \psi^{\alpha}_a \psi^{a \beta} ,
\end{equation}
with the new fermionic variables $\bar \psi^{\alpha}_a$  and
$ \psi^{a \beta}$ connected with the old ones via the tetrad
$e^\alpha_i$ ($e^\alpha_i e^\beta_j \eta_{\alpha \beta} = g_{ij}$)
\begin{equation} \label{TETRAD}
\bar \psi^{\alpha}_a = e^\alpha_i \bar \psi^{i}_a \quad
\hbox{and} \quad
 \psi^{\alpha}_a = e^\alpha_i  \psi^{i}_a ,
\end{equation}
and  $\omega_{i \alpha \beta}$ in (\ref{IMP}) is the corresponding spin
connection. Therefore,
the quantum supercharges (\ref{QUANT1}) -- (\ref{QUANT2}) are mutually
Hermitean conjugated and the resulting quantum Hamiltonian $H_{quant.}$
is a Hermitean self -- adjoint operator as well.

As a result,  equations (\ref{OOO}) -- (\ref{TETRAD})
completely describe  the general formalism of the $N=4$
SUSY $D$~-- dimensional quantum mechanics, and this provides the basis
for the analysis of its main properties. \\

\setcounter{equation}0
\section{Partial SUSY Breaking}

Let us investigate in detail the question of  partial supersymmetry
breakdown in the framework of the constructed $N=4$ SUSY QM in
an arbitrary $D$ number of dimensions. As it has been mentioned in the
introduction, the problem of partially broken supersymmetry is very
important for applications in supergravity, superstring theories and
in the $M$~-- theory as well, and the
supersymmetric quantum mechanics turns out
to be an adequate mighty tool for investigating of the corresponding
problems in supersymmetric field theories.

We shall see that in contrast with the one -- dimensional $N=4$
SUSY QM, the multidimensional one provides also
 possibilities when either only one
quarter of all supersymmetries is exact (for $D \geq 2$),
or one quarter
of all supersymmetries is broken (for $D \geq 3$).

In order to study partial SUSY breaking it is convenient
to introduce a new set of real -- valued supercharges
\begin{equation}
S^a = Q_a + \bar Q^a ,
\end{equation}
\begin{equation}
T^a = i(Q_a - \bar Q^a) .
\end{equation}
In the above equations  the $SU(2)$ covariance is obviously
damaged. This is the price we pay for passing to the real--valued
supercharges. However, for a further discussion the loss of the
covariance does not cause any problems. The label ``a" has now to be
considered
 just as the number of  supercharges denoted by $S$ and $T$.

The new supercharges form the following $N=4$ superalgebra with
the central charges
\begin{eqnarray} \label{ALGEBRA1}
\{ S^a, S^b \}&=& H (\delta ^a_b +\delta^b_a)
+  (\lambda^a_{bi} +\lambda^b_{ai})m^i +
(\lambda _{abi} n^i - \lambda^{ab}_i \bar n^i) ,  \\  \label{ALGEBRA2}
\{ T^a, T^b \}&=& H (\delta ^a_b +\delta^b_a)
+  (\lambda^a_{bi} +\lambda^b_{ai})m^i -
(\lambda _{abi} n^i - \lambda^{ab}_i \bar n^i) , \\  \label{ALGEBRA3}
\{ S^a, T^b \}&=&i (\lambda^a_{bi}  - \lambda^b_{ai})m^i +
i(\lambda _{abi} n^i + \lambda^{ab}_i \bar n^i) ,
\end{eqnarray}
where $\lambda^{a}_{bi}=(\sigma_I)^a_b \Lambda^I_i$ and
$\Lambda^I_i$ are real parameters.

The algebra (\ref{ALGEBRA1}) -- (\ref{ALGEBRA3}) is still
nondiagonal. However, some particular choices of the
constant parameters $m^i, n^i$ and $\Lambda_{i}^I$ bring
the algebra to the standard form, i.e., to the form
when the right -- hand side of (\ref{ALGEBRA3}) vanishes and
the right -- hand sides of (\ref{ALGEBRA1}) and (\ref{ALGEBRA2})
are diagonal with respect to the indices ``a" and ``b".

Now we consider several cases separately.

\subsection{ Four supersymmetries exact / Four supersymmetries broken}

If we put equal to zero all central charges, appearing
in the algebra, then no partial breakdown of supersymmetry
is possible. In this case, all supersymmetries are exact,
if the energy of the ground state is zero; otherwise all of them are
broken. This statement is obviously independent of the number of
dimensions $D$.

\subsection{Two supersymmetries exact}

The case of  partial supersymmetry breakdown,
 when the half of  supersymmetries are exact, have been
considered earlier \cite{IKP} in the framework of one-dimensional
$N=4$ SUSY QM,  but we shall describe it  for completeness as well.
Consider the one -- dimensional ($D=1$) $N=4$ SUSY QM and put all
the constants entering into the right -- hand sides
of (\ref{ALGEBRA1}) -- (\ref{ALGEBRA3})
equal to zero, except
\begin{equation} \label{SET1}
m^1 \quad \hbox{and} \quad \Lambda^3_1.
\end{equation}
Then, the algebra (\ref{ALGEBRA1}) -- (\ref{ALGEBRA3}) takes the form
\begin{eqnarray} \nonumber
\{ S^1 , S^1 \}&=&2H + 2 m^1  \Lambda^3_1 , \\ \nonumber
\{ S^2 , S^2 \}&=&2H - 2 m^1  \Lambda^3_1 , \\ \nonumber
\{ T^1 , T^1 \}&=&2H + 2 m^1  \Lambda^3_1 , \\ \label{TWO}
\{ T^2 , T^2 \}&=&2H - 2 m^1  \Lambda^3_1 .
\end{eqnarray}
It means that if the energy of the ground state is equal
to $m^1  \Lambda^3_1$ and the last-mentioned product is positive,
then $S^2$ and $T^2$ supersymmetries are exact, while the other
two are broken. If $m^1  \Lambda^3_1$ is negative, then
$S^1$ and $T^1$ supersymmetries are exact provided the energy of the
ground state is equal to  $- m^1  \Lambda^3_1$.

\subsection{ One supersymmetry exact}
The case of the three -- quarters breakdown of supersymmetry
is possible if the dimension of $N=4$ SUSY QM
is at least two ($D \geq 2$).
Indeed, for $D=2$ let us keep  the following set of parameters nonvanished:
\begin{equation} \label{SET2}
\Lambda^3_1, \Lambda^1_2, m^1 \quad \hbox{and} \quad Re (n^2) \equiv N^2 .
\end{equation}
Then, one obtains
\begin{eqnarray}  \nonumber
\{ S^1 , S^1 \}&=&2H + 2 m^1  \Lambda^3_1  - 2 \Lambda^1_2 N^2 , \\
\nonumber
\{ S^2 , S^2 \}&=&2H - 2 m^1  \Lambda^3_1 + 2 \Lambda^1_2 N^2 , \\
\nonumber
\{ T^1 , T^1 \}&=&2H + 2 m^1  \Lambda^3_1 + 2 \Lambda^1_2 N^2 , \\
\label{ONE}
\{ T^2 , T^2 \}&=&2H - 2 m^1  \Lambda^3_1 - 2 \Lambda^1_2 N^2 .
\end{eqnarray}
{}Further choice
\begin{equation} \label{CONDITION2}
m^1  \Lambda^3_1 =  \Lambda^1_2 N^2
\end{equation}
leads to the  case when only $T^2$ supersymmetry is exact, while
all others are broken if the energy  of the ground state is
equal to $2m^1  \Lambda^3_1$,  and $m^1  \Lambda^3_1 > 0$.
 If $m^1  \Lambda^3_1$ is negative, then  $T^1$ is exact
provided the energy of the ground state is equal to
$- m^1  \Lambda^3_1$.

\subsection{ Three supersymmetries exact}

The situation of the one -- quarter
breakdown of supersymmetry  can exist, if we add to the consideration
one more dimension, i.e., consider the
three -- dimensional $D=3$ $N=4$ supersymmetric quantum mechanics.

  Keeping the following
set of the parameters nonvanished
\begin{equation} \label{SET3}
\Lambda^3_1, \Lambda^1_2, \Lambda^2_3, m^1,  N^2  \quad
\hbox{and} \quad Im (n^3)
\equiv M^3,
\end{equation}
we have
\begin{eqnarray}  \nonumber
\{ S^1 , S^1 \}&=&2H + 2 m^1  \Lambda^3_1  - 2 \Lambda^1_2 N^2
- 2 \Lambda^2_3 M^3 , \\ \nonumber
\{ S^2 , S^2 \}&=&2H - 2 m^1  \Lambda^3_1 + 2 \Lambda^1_2 N^2
- 2 \Lambda^2_3 M^3 , \\ \nonumber
\{ T^1 , T^1 \}&=&2H + 2 m^1  \Lambda^3_1 + 2 \Lambda^1_2 N^2
+ 2 \Lambda^2_3 M^3 , \\  \label{THREE}
\{ T^2 , T^2 \}&=&2H - 2 m^1  \Lambda^3_1 - 2 \Lambda^1_2 N^2
+ 2 \Lambda^2_3 M^3 .
\end{eqnarray}
If
\begin{equation} \label{CONDITION31}
m^1  \Lambda^3_1  =  \Lambda^1_2 N^2,
\end{equation}
\begin{equation} \label{CONDITION32}
 \Lambda^1_2 N^2 = - \Lambda ^2_3 M^3,
\end{equation}
and
\begin{equation} \label{CONDITION33}
m^1  \Lambda^3_1 < 0,
\end{equation}
then $T^2$ supersymmetry is broken, while all others are exact
under the condition that the energy of the ground state is equal
to $m^1  \Lambda^3_1$. If the last-mentioned product is positive, then
$T^2$ supersymmetry is exact, while all others are broken
provided that the energy of the ground state is
$3 m^1  \Lambda^3_1$ and we arrive at the three-dimensional
generalization of the  case {\bf C}.

Obviously, when considering the three -- dimensional $N=4$ SUSY QM,
one can either keep  the parameters (\ref{SET2}) under the condition
(\ref{CONDITION2}), or the parameters
(\ref{SET1}), or put all of them equal to zero and, therefore,
obtain all particular cases of spontaneous breakdown of supersymmetry
discussed earlier.
It is also obvious that all these cases can be obtained from the
higher dimensional ($D \geq 3$) $N=4$ supersymmetric quantum mechanics.

To summarize this section one should note that according to the given
general analysis of partial SUSY breaking in the $N=4$ multidimensional
SUSY QM, there exist possibilities of constructing the models with
$\frac{1}{4}$, $\frac{1}{2}$ and $\frac{3}{4}$ supersymmetries
unbroken, as well as models with totally broken or totally
unbroken supersymmetries. However, the answer to the question
which  of these possibilities can be realized for the considered
system, crucially depends  on the form of the chosen superpotential
and on the imposed boundary conditions of the quantum mechanical
problem.
 \\

\setcounter{equation}0\section{Explicit Example}

{}For  better illustration of the ideas of the previous
section it is useful to consider a particular choice
of the superpotential $A(\Phi^i)$. As it has been mentioned before,
to consider all possible cases of  partial supersymmetry
breakdown, the minimal amount of the superfields needed is three.
Therefore, let us take three superfields of the type
(\ref{SF}) and choose the
 constants $m^i, n^i, \bar n^i$ and  $\Lambda^I_{i}$
in accordance with   expressions
(\ref{SET1}), (\ref{SET2}) and (\ref{SET3}).

A simple and at the same time interesting example is
 the case when the superpotential is the direct sum
of terms, each being a function of only one superfield.
This gives the possibility of the considerable
 simplification of the classical
and quantum Hamiltonians,  and the supercharges as well
\cite{IKP}.
We choose the explicit form of the superpotential as
\begin{equation} \label{AAA}
A(\Phi^i) = \Phi^i ln \Phi^i , \quad i = 1,2,3
\end{equation}
and consider the physical bosonic components of the
 superfields $\Phi^i$ as  functions of the new
variables $x^i$, namely,
\begin{equation}  \label{BBB}
\phi^i = {(x^i)}^2 .
\end{equation}
 Making the following redefinition of the fermionic variables
\begin{equation} \label{CCC}
\xi^{ai} = \psi^{ai}\sqrt{2\frac{\partial^2 A}{{(\partial \phi^{i})}^2}} ,
\quad
\bar \xi^{i}_a = \bar \psi^{i}_a
\sqrt{2\frac{\partial^2 A}{{(\partial \phi^{i})}^2}} ,
\end{equation}
where no summation over the repeated indices is assumed,
one obtains the canonical commutation relations between fermions
\begin{equation} \label{DDD}
\{ \xi^{ai} , \bar \xi^j_b \} = \delta^a_b \delta^{ij} .
\end{equation}
Inserting   expressions
(\ref{AAA}), (\ref{BBB}) and (\ref{CCC}) into  (\ref{OPERATOR}),
one obtains the three -- dimensional
superconformal $N=4$ quantum mechanics \cite{IKL} with
\begin{equation} \label{CONF}
H_{quant.} = H^1 + H^2 + H^3 ,
\end{equation}
i.e., as it could
be concluded from the fact that the superpotential is diagonal with
respect to the superfields considered,
the total Hamiltonian is also a direct sum of three Hamiltonians,
each of them containing  the bosonic and fermionic operators of only
one type. The explicit form of the Hamiltonians $H^i$, $(i =1,2,3.)$
is
\begin{eqnarray} \nonumber
H^1&=& - \frac{1}{8} \frac{d^2}{{(d x^1)}^2}
+ \frac{1}{4}\Lambda^3_1 (\sigma_3)_a^b \bar \xi^1_b \xi^{a1}
+\frac{1}{8}{(\Lambda^3_1)}^2 (x^1)^2\\ \label{HAM1}
&& +\frac{1}{(x^1)^2}( 2 (m^1)^2 + \frac{3}{32} - m^1
(\bar \xi^1_a \xi^{a1}
-1) - \frac{1}{4} \bar \xi^1_a \xi^{a1} +
\frac{1}{8}(\bar \xi^1_a \xi^{a1})(\bar \xi^1_b \xi^{b1})) ,
\end{eqnarray}

\begin{eqnarray} \nonumber
H^2&=& - \frac{1}{8} \frac{d^2}{{(d x^2)}^2}
+ \frac{1}{4}\Lambda^1_2 (\sigma_1)_a^b \bar \xi^2_b \xi^{a2}
+\frac{1}{8}{(\Lambda^1_2)}^2 (x^2)^2\\ \label{HAM2}
&& +\frac{1}{(x^2)^2}( 2 (N^2)^2 + \frac{3}{32} -
\frac{1}{2}N^2 ( \xi^{a2} \xi_a^{2} + \bar \xi^2_a \bar \xi^{a2})
 - \frac{1}{4} \bar \xi^2_a \xi^{a2} +
\frac{1}{8}(\bar \xi^2_a \xi^{a2})(\bar \xi^2_b \xi^{b2})) ,
\end{eqnarray}

\begin{eqnarray} \nonumber
H^3&=& - \frac{1}{8} \frac{d^2}{{(d x^3)}^2}
+ \frac{1}{4}\Lambda^2_3 (\sigma_2)_a^b \bar \xi^3_b \xi^{a3}
+\frac{1}{8}{(\Lambda^2_3)}^2 (x^3)^2\\ \label{HAM3}
&& +\frac{1}{(x^3)^2}( 2 (M^3)^2 + \frac{3}{32} -
\frac{i}{2} M^3 ( \xi^{a3} \xi^{3}_a  -
\bar \xi^3_a \bar \xi^{a3}) - \frac{1}{4} \bar \xi^3_a \xi^{a3} +
\frac{1}{8}(\bar \xi^3_a \xi^{a3})(\bar \xi^3_b \xi^{b3})) .
\end{eqnarray}
The next step is to find the energy spectrum of the quantum
Hamiltonian (\ref{CONF}).

Since the bosonic and fermionic variables of each type
are completely separated,  the eigenfunctions
of the Hamiltonian  (\ref{CONF}) is a direct product of the
eigenfunctions of the Hamiltonians  (\ref{HAM1}) - (\ref{HAM3})
and the total energy is just a sum of the energies
corresponding to the Hamiltonians
$H^i$.

Let us find the energy spectrum
of the Hamiltonian $H^1$.
 Consider the general state in the ``reduced" Fock space
spanned by the fermionic creation and annihilation operators
$\bar \xi_a^1$ and $\xi^{a1}$ obeying the anticommutation relations
(\ref{DDD})  with $i=1$:
\begin{equation} \label{STATE}
|\rho\rangle = X_1(x^1)|0\rangle + Y_1^a(x^1) \bar \xi^1_a|0\rangle +
Z_1(x^1)\bar \xi^1_a \bar \xi^{a1}|0\rangle .
\end{equation}
The operator $H^1$, acting on the state vector (\ref{STATE}), gives
the following four Shr\"odinger equations for
the unknown functions $X_1(x^1)$, $Y_1^a(x^1)$
and $Z_1(x^1)$:
\begin{equation}
(-\frac{1}{2} \frac{d^2}{{(d x^1)}^2}
+\frac{1}{2}{(\Lambda^3_1)}^2 (x^1)^2
+\frac{1}{(x^1)^2}( 8 (m^1)^2 + 4 m^1 + \frac{3}{8}))X_1(x^1)
=4E^1_I X_1(x^1) ,
\end{equation}

\begin{equation}
(-\frac{1}{2} \frac{d^2}{{(d x^1)}^2} + \Lambda^3_1
+\frac{1}{2}{(\Lambda^3_1)}^2 (x^1)^2
+\frac{1}{(x^1)^2}( 8 (m^1)^2 - \frac{1}{8}))Y^1_1(x^1)
=4E^1_{II}Y_1^1(x^1) ,
\end{equation}

\begin{equation}
(-\frac{1}{2} \frac{d^2}{{(d x^1)}^2} - \Lambda^3_1
+\frac{1}{2}{(\Lambda^3_1)}^2 (x^1)^2
+\frac{1}{(x^1)^2}( 8 (m^1)^2 - \frac{1}{8}))Y_1^2(x^1)
=4E^1_{III} Y_1^2(x^1) ,
\end{equation}

\begin{equation}
(-\frac{1}{2} \frac{d^2}{{(d x^1)}^2}
+\frac{1}{2}{(\Lambda^3_1)}^2 (x^1)^2
+\frac{1}{(x^1)^2}( 8 (m^1)^2 - 4 m^1 + \frac{3}{8}))Z_1(x^1)
=4E^1_{IV} Z_1(x^1) .
\end{equation}
The wave functions and the energy spectrum of the Hamiltonian
of the type
\begin{equation} \label{PER}
{\cal H} = -\frac{1}{2}\frac{d}{dx^2} + \frac{1}{2} x^2 + g \frac{1}{x^2} ,
\end{equation}
have been investigated in detail for the nonsupersymmetric theory
\cite{LATH} -- \cite{DAS}
and in the framework of the $N=2$ supersymmetric quantum mechanics
\cite{DAS} -- \cite{FR} as well.
The most detailed and complete study has been done by
Das and Pernice \cite{DAS} where the eigenfunctions and  energy
spectrum of the Hamiltonian of the type (\ref{PER}) were found
after appropriate regularization of the potential and superpotential,
depending on whether one considers nonsupersymmetric or $N=2$ supersymmetric
problem. However, as it can be seen from ({\ref{AAA}) and (\ref{BBB})
the superpotential in our $N=4$ case for the Hamiltonian
with the $\frac{1}{x^2}$ term in the potential energy  is regular
in contrast with the case of $N=2$ supersymmetric
quantum mechanics and,
therefore, we  use the results of  \cite{DAS} which
are obtained after the regularization of the  potential, but not
of the superpotental.

{}For the problem considered one obtains
(we take the value of the parameter $\Lambda^3_1$ without loss
of generality to be equal to $+1$).

{}For $m^1 < - \frac{1}{4}$,
\begin{eqnarray} \nonumber
4E^{1}_I&=& 2k^1_I -4m^1  , \\ \nonumber
4E^{1}_{II}&=& 2k^1_{II} - 4m^1 + 2 , \\ \nonumber
4E^{1}_{III}&=& 2k^1_{III } - 4m^1 ,  \\  \label{SPECTR1}
4E^{1}_{IV}&=& 2k^1_{IV} - 4m^1 + 2 ,
\end{eqnarray}
where $k^A_M = 0,1,2,...,$, $(A= 1,2,3,4)$ and $(M=I,II,III,IV)$.
Each energy level $E^A_M$ corresponds to a couple
(even and odd) of wave functions and, therefore, is doubly degenerate.
The minimal energy corresponds
 to the minima of $E^1_I$ and $E^1_{III} $ for $k^1_I = k^1_{III} = 0$
and equals  $-m^1$. Let us denote the corresponding
states by $\pi^{ 1\pm}_I $ and $\pi^{ 1\pm}_{III} $.

{}For $-\frac{1}{4} < m^1 < 0$ one has
\begin{eqnarray} \nonumber
4E^{1}_I&=& 2k^1_I + 4m^1 + 2 , \\ \nonumber
4E^{1}_{II}&=& 2k^1_{II} - 4m^1 + 2 , \\ \nonumber
4E^{1}_{III}&=& 2k^1_{III } - 4m^1 ,   \\  \label{SPECTR2}
4E^{1}_{IV}&=& 2k^1_{IV} - 4m^1 + 2 .
\end{eqnarray}

The minimal energy corresponds
 to the minimum of $E^1_{III}$ for $k^1_{III} = 0$
and equals  $-m^1$.  We denote the corresponding
ground states by $\pi^{ 1\pm}_{III} $.

{}For $0 < m^1 < \frac{1}{4}$ one has
\begin{eqnarray} \nonumber
4E^{1}_I&=& 2k^1_I + 4m^1 + 2 , \\ \nonumber
4E^{1}_{II}&=& 2k^1_{II} + 4m^1 + 2 , \\ \nonumber
4E^{1}_{III}&=& 2k^1_{III } + 4m^1 ,   \\  \label{SPECTR3}
4E^{1}_{IV}&=& 2k^1_{IV} - 4m^1 + 2 .
\end{eqnarray}

The minimal energy is $m^1$ for  $k^1_{III} = 0$, and the
corresponding ground state is again  $\pi^{ 1\pm}_{III} $.

 Finally,
for $m^1 >  \frac{1}{4}$:
\begin{eqnarray} \nonumber
4E^{1}_I&=& 2k^1_I + 4m^1 + 2 ,  \\ \nonumber
4E^{1}_{II}&=& 2k^1_{II} + 4m^1 + 2 , \\ \nonumber
4E^{1}_{III}&=& 2k^1_{III } + 4m^1 ,   \\  \label{SPECTR4}
4E^{1}_{IV}&=& 2k^1_{IV} + 4m^1 ,
\end{eqnarray}

The minimal energy is $m^1$ for  $k^1_{III} = k^1_{IV} = 0$,
and the corresponding ground states are $\pi^{ 1\pm}_{III} $ and
$\pi^{ 1\pm}_{IV}$.

The points $\pm \frac{1}{4}$, and $0$ are the branching points.
 When  $m^1$ gets these values,  the corresponding
energies and wave functions of the system in  the regions of the parameter,
divided
by these points,  just coincide.

If we also choose $\Lambda^1_2=\Lambda^2_3 = 1$,
the energy spectra of the Hamiltonians $H^2$ and $H^3$ are absolutely
the same as in (\ref{SPECTR1}) -- (\ref{SPECTR4}).
The only difference is that the parameter $m^1$ should
be replaced by $N^2$ or $M^3$ respectively.
However, the eigenfunctions, corresponding to
$E^{2}_{I}$ and $E^{2}_{IV}$, are  linear combinations
of the states of zero and two fermionic sectors since
the fermionic number operator $\bar \xi^2_a \xi^{a2}$ does not
commute with the Hamiltonian $H^2$.
The energies $E^{2}_{II}$ and $E^{2}_{III}$ are also  linear
combinations
of  both the states of one fermionic sector  because the matrix
$(\sigma_1)^b_a$ is not diagonal. An analogous situation takes
place for the Hamiltonian $H^3$.

Now we are in a position to  describe partial supersymmetry breaking
following the lines of the previous section.

{}First, let us consider the one -- dimensional  case with $m^1$
 equal to zero. As it has been mentioned above, zero value of $m^1$ is the
branching point and, therefore, the energy spectra (\ref{SPECTR2}) and
(\ref{SPECTR3}) as well as the wavefunctions in these regions completely
coincide. Therefore, one has the couple of supersymmetric
ground states $\pi^{ 1\pm}_{III}$  and all supersymmetries are exact.

As it has been mentioned in the previous Section, in order to describe
the halfbreaking of supersymmetry it is enough to consider
only the spectrum of the Hamiltonian $H^1$.
Inserting the corresponding eigenvalues
of the operator $H^1$ for each range of the parameter
$m^1$ into equations (\ref{TWO}), one obtains that half
of supersymmetries are always broken.

Considering the spectra of the Hamiltonians
$H^1$ and $H^2$, one can obtain the three -- quarter breakdown of
supersymmetry.
Indeed, from  equations  (\ref{ONE}), (\ref{CONDITION2})
and (\ref{SPECTR1}) -- (\ref{SPECTR4}), one can conclude
that either $T^1$ or $T^2$ supersymmetries are exact depending
on the range of the parameter $m^1$. The corresponding ground state
wavefunctions obviously are
\\ for $m^1 < - \frac{1}{4}$
\begin{equation}
\pi_I^{1 \pm} \times \pi_I^{2 \pm} , \quad
\pi_I^{1 \pm} \times \pi_{III}^{2 \pm} , \quad
\pi_{III}^{1 \pm} \times \pi_I^{2 \pm} , \quad
\pi_{III}^{1 \pm} \times \pi_{III}^{2 \pm} ,
\end{equation}

for $-\frac{1}{4}< m^1 < 0$:
\begin{equation}
\pi_{III}^{1 \pm} \times \pi_{III}^{2 \pm} ,
\end{equation}

 for $0< m^1 <  \frac{1}{4}$
\begin{equation}
\pi_{III}^{1 \pm} \times \pi_{III}^{2 \pm} ,
\end{equation}

 for $m^1 > \frac{1}{4}$:
\begin{equation}
\pi_{III}^{1 \pm} \times \pi_{III}^{2 \pm} , \quad
\pi_{III}^{1 \pm} \times \pi_{IV}^{2 \pm} , \quad
\pi_{IV}^{1 \pm} \times \pi_{III}^{2 \pm} , \quad
\pi_{IV}^{1 \pm} \times \pi_{IV}^{2 \pm} .
\end{equation}

In order to study the possibility of the one -- quarter breakdown
of supersymmetry, one has to consider the three -- dimensional case,
i.e., the spectra and the wavefunctions of the Hamiltonians
$H^1$, $H^2$ and $H^3$. Using  equations
(\ref{THREE}), (\ref{CONDITION31}) -- (\ref{CONDITION33}) and
(\ref{SPECTR1}) -- (\ref{SPECTR4}) one can conclude that for the
considered
model the one -- quarter supersymmetry breakdown is impossible since
the energy of the ground state equals  $3m^1$ rather than  $m^1$,
as it is required for the annihilation of the ground
state by the operators $S^1$, $S^2$
and $T^1$. This obviously does not mean that
one-quarter supersymmetry breakdown is impossible in principle;
it means instead that this effect is impossible for the
simple model we considered.

Indeed, let us consider the same three -- dimensional
problem, but restricting ourselves
to non -- negative values of coordinate $x^1$, i.e.,
$x^1 \geq 0$.

The  spectrum of ${\cal H}$ (\ref{PER}), when $x$ belongs
to the non -- negative half-axis is slightly
different \cite{LATH}\footnote{In fact, as it has been
recently shown by A. Das and S. Pernice \cite{DAS}, the
energy spectrum, obtained in \cite{LATH} is correct if one considers the
problem only on the half-axis.}
and it opens a possibility of constructing the ground state which is
invariant under three
unbroken supersymmetries. According to  \cite{LATH}, we have
\begin{equation}
E_k^{(\pm \alpha)} = 2k \pm \alpha + 1 ,
\end{equation}
where $\alpha$ is given by
\begin{equation}
\alpha = + \frac{1}{2} \sqrt {1 + 8g} ,
\end{equation}
and $k$ is the non -- negative integer.
If $\alpha \geq 1$, then the energies
 $E_k^{(-\alpha)}$ must be excluded from the spectrum
since the corresponding wave functions are no longer normalizable.
Otherwise one has to consider both sets of solutions.
Applying these results to the problem
under consideration, and putting again
$\Lambda^3_1 = \Lambda^1_2 = \Lambda^2_3 =1$,
 one obtains for $H^1$
\begin{eqnarray} \nonumber
\alpha^1_I&=&|4m^1 + 1| , \\ \nonumber
\alpha^1_{II}&=&|4m^1| ,  \\ \nonumber
\alpha^1_{III}&=&|4m^1| ,  \\ \label{ALPHA}
\alpha^1_{IV}&=&|4m^1 - 1| .
\end{eqnarray}
And, therefore, the energy spectra have the form
\begin{eqnarray} \nonumber
4E^{1, (\pm)}_I&=& 2k^1_I \pm |4m^1 +1| + 1 , \\ \nonumber
4E^{1, (\pm)}_{II}&=& 2k^1_{II} \pm |4m^1| + 2 , \\ \nonumber
4E^{1, (\pm)}_{III}&=& 2k^1_{III } \pm |4m^1|  , \\ \label{SPECTR}
4E^{1, (\pm)}_{IV}&=& 2k^1_{IV} \pm |4m^1 - 1| + 1 ,
\end{eqnarray}
 Both the signs before the second terms have to be taken
for $E_I$ if $ - \frac{1}{2} < m^1 < 0$ ;
for $E_{II}$ and for $E_{III}$ if $ - \frac{1}{4} < m^1 < \frac{1}{4}$ ;
for $E_{IV}$ if $0 < m^1 < \frac{1}{2}$.
Let us further restrict the value of the parameter so
that it belongs  to the open
interval $-\frac{1}{4} < m^1 < 0$. Then due to the equations
(\ref{CONDITION31}) -- (\ref{CONDITION33}),
(\ref{SPECTR2}) -- (\ref{SPECTR3}),
 and (\ref{SPECTR})
the minimal energy of the system
with $k^1_{III} = k^2_{III} = k^3_{III}=0$ is
\begin{equation}
E_{min.}=E_{III}^{1,-} + E_{III}^{2,\pm} + E_{III}^{3,\pm}  = - m^1,
\end{equation}
and according to (\ref{THREE}) we have the supersymmetric ground states
with three supersymmetries being unbroken.

In this section we have considered quite schematically
the one-, two- and three-dimensional $N=4$ supersymmetric versions of
the quantum oscillator
with the additional $\frac{1}{x^2}$ term in the potential energy.
However, we believe that even this simple analysis
gives a good illustration of all possible cases of the partial
supersymmetry breakdown in the
multidimensional $N=4$ SUSY QM. One should also stress the
crucial meaning of the boundary conditions in the question of
 partial supersymmetry breakdown, as it has been shown for the case of
 one -- quarter supersymmetry breakdown in the considered example.

\setcounter{equation}0\section{Discussion}
In this paper, we have described the general formalism of the
multidimensional $N=4$ supersymmetric quantum mechanics and studied
 various possibilities of partial supersymmetry breaking illustrating
them by the exactly solvable example.

 However,  several
questions, which seem to be of  particular importance are left still
opened.
Indeed, it would be interesting to investigate other
possibilities of changing  the bosonic end fermionic variables,
namely, for the cases, when in contrast with (\ref{AAA}) and (\ref{BBB})
the superpotential $A(\Phi^i)$ is not a direct sum of the terms, each
containing only one superfield $\Phi^i$ and when the bosonic
components of these superfields depend on several variables $x^i$.
The detailed study of this problem can lead to  possible $N=4$
supersymmetrization and quantization of various pure bosonic integrable
systems such as $n$~-particle Calogero and Calogero -- Moser models,
which are related to the RN black hole quantum mechanics and to
$D=2$ SYM theory \cite{GT}.
This approach can also  answer the question about the general
class of  potentials which lead to  superconformal $N=4$ theories
in higher dimensions.

Another topic, which is left uncovered in this paper, is
the possible application of the constructed multidimensional $N=4$ SUSY QM
to the problems of quantum cosmology.  Possible embedding of
pure bosonic effective Lagrangians, derived from the homogeneous
cosmological models to  $N=4$ SUSY QM can shed a new light on the
old problems of boundary conditions and spontaneous SUSY breaking
in quantum cosmology which have recently been investigated
 in the framework
of $N=2$ supersymmetric sigma -- model approach \cite{BG} -- \cite{DTT}.

All these questions are now under intensive study  and will be reported
elsewhere.

\setcounter{equation}0\section{Acknowledgments}
We would like to thank  E. A. Ivanov for the helpful and stimulating
discussions and A. V. Gladyshev and C. Sochichiu for some comments.
 Work of A. P. was supported in part by INTAS Grant 96-0538
and by the
Russian Foundation for Basic Research,
Grant 99-02-18417.
Work of M. T. was supported in part by INTAS Grant 96-0308.
J. J. R. would like to thank CONACyT for the  support under the
program: Estancias Posdoctorales en el Extranjero, and Bogoliubov
Laboratory of JINR for hospitality. \\

\end{document}